\documentclass[onefignum,onetabnum]{siamart251216}
\usepackage[T1]{fontenc}
\usepackage{amssymb,mathtools}
\usepackage{array,booktabs}

\newsiamthm{conjecture}{Conjecture}
\newsiamthm{pr}{Property}
\newsiamremark{remark}{Remark}
\newcommand{\C}{\mathcal{C}}
\newcommand{\F}{\mathbb{F}}

\newcommand{\PGRM}{\operatorname{PGRM}}
\newcommand{\wt}{\operatorname{wt}}
\newcommand{\wtq}{\operatorname{wt}_q}
\newcommand{\cl}{\operatorname{cl}}

\newcommand{\Tr}{\operatorname{Tr}}

\title{Exact Minimum Distances of Primitive Narrow-Sense BCH Codes: At and Beyond the Bose Distance}
\author{Run Zheng\thanks{Department of Mathematics,
The Hong Kong University of Science and Technology, Hong Kong, China
(\email{zhengrung@ust.hk}).}}

\begin{document}
\maketitle
\begin{abstract}
 We determine the exact minimum
  distances of two families of primitive narrow-sense BCH 
  codes of length $q^m-1$, where $q$ is a prime power 
  and $m\geq 4$. For $1\leq i\leq\lfloor m/2\rfloor$, codes 
  with designed distance $q^m-q^{m-1}-q^{m-1-i}-1$
   have minimum distance and   Bose distance both equal
    to their designed distance.  Furthermore, for integers $\ell, i, s$ 
     satisfying $3\leq\ell<m-1$ and 
     $\lfloor m/(\ell+1)\rfloor\leq i\leq s<\lfloor(m-1)/\ell\rfloor$, codes with designed distance $q^m-q^{m-1}-q^{m-1-i}-q^s-1$ have Bose distance equal to their designed distance and minimum distance $q^m-q^{m-1}-q^{m-1-i}-1$. Consequently, the gap between the minimum distance and the
Bose distance in this family is exactly $q^s$ and grows
without bound as $m$ increases, with fixed $\ell$.
The proofs proceed by constructing good zero-sets from
quadratic forms for the first family, and by combining
upper bounds provided by the first family with lower bounds
derived from the divisibility of codeword weights in   punctured generalized Reed--Muller codes  for the second.
We also derive explicit 
dimension formulas by counting cyclotomic cosets.
\end{abstract}


\section{Introduction}\label{section1}
Throughout this paper, let $q$ be a prime power, let $m\geq1$ be an
integer, and let $n=q^m-1$. Denote by $\C_{(q,m,\delta)}$ the $q$-ary
primitive narrow-sense Bose--Chaudhuri--Hocquenghem (BCH) code
\cite{bose,hocquenghem} of length $n$ with designed distance $\delta$,
where $2\leq\delta\leq n$. It may happen that  distinct designed distances 
define the same code.  The largest designed distance defining
$\C_{(q,m,\delta)}$ is called its \emph{Bose distance}, denoted by
$d_B(\C_{(q,m,\delta)})$. By the BCH bound, the minimum distance
$d(\C_{(q,m,\delta)})$ satisfies
\begin{equation}\label{equation1}
d\bigl(\C_{(q,m,\delta)}\bigr)
\geq d_B\bigl(\C_{(q,m,\delta)}\bigr)
\geq\delta.
\end{equation}
When the code is clear from the context, we simply write $d$ and $d_B$
for its minimum distance and Bose distance, respectively. Although
primitive narrow-sense BCH codes have a simple algebraic definition and are widely used,
their parameters are known only in very limited cases.  Determining their dimensions and Bose distances is already a challenging task that has attracted extensive research \cite{CherchemEtAl2020,DingDuZhou2015,GanEtAl2024,LiuDingLi2017,MoulouaNajmeddine2023,PangZhuKai2021,YueFeng2000,ZhengSzeHuang2025}. Finding their exact minimum distances remains an even more difficult problem  \cite{Charpin1998,DingLi2024}.

Almost all known infinite families of primitive narrow-sense BCH codes
whose exact minimum distances have been determined satisfy $d=d_B$ (or
$d=\delta$). Since \eqref{equation1} already gives the lower
bound $d\geq d_B$, proving $d=d_B$ (or $d=\delta$) only requires
establishing the existence of a codeword of weight $d_B$ (or $\delta$).
Classical results include Peterson's family with designed distance
$\delta=q^t-1$, where $1\leq t\leq m-1$ \cite{Peterson1969}.
Further families have been obtained using connections with
Reed--Muller codes \cite{Berlekamp1970,Ding2015,KasamiLin1972},
idempotents \cite{AugotSendrier1994,ShanyBerman2025}, and trace
representations together with quadratic forms
\cite{DingFanZhou2017,Li2017}.
Cohen \cite{Cohen1997} derived exact minimum distances from
covering-radius estimates, and Noguchi et al.\
\cite{NoguchiEtAl2021} obtained families with explicitly determined
minimum distances proportional to the code length.
More recent constructions use locator polynomials
\cite{ChenEtAl2026}, a related criterion for Goppa codes
\cite{ChenEtAlGoppaBCH2026}, and $\F_q$-good zero-sets
\cite{zhenggood}.

Determining the exact minimum distance becomes substantially harder
when $d>d_B$. In this case, the lower bound \eqref{equation1}
is no longer tight, so determining $d$ requires not only establishing
the existence of a codeword of weight $d$, but also ruling out all
nonzero codewords of any weight from $d_B$ to $d-1$.
 Consequently, exact results in the regime $d>d_B$ are scarce.

 Early work of Kasami and Tokura \cite{KasamiTokura1969} gave some binary
examples with $d>d_B$. Based on numerical computations and known
results, it was widely conjectured that $d-d_B\leq4$; this was
recorded by Charpin \cite{Charpin1998} and was subsequently referred
to as Charpin's conjecture. Recently, Zheng et al.\ \cite{ZhengEtAl2026} disproved this conjecture  by showing that, for any prime power $q$, the gap $d - d_B$ can be arbitrarily large for $q$-ary primitive narrow-sense BCH codes. Moreover, they explicitly determined the exact minimum distances in the binary case. To the best of our
knowledge, this provides the first infinite family of binary primitive
narrow-sense BCH codes satisfying $d>d_B$ whose exact minimum distance
has been determined, whereas no such nonbinary family has previously
been known.

In this paper, we determine the exact minimum distance of an infinite
family of $q$-ary primitive narrow-sense BCH codes satisfying $d>d_B$ for
every prime power $q$, and we also obtain its Bose distance and
dimension. Our first main result is the following theorem.
\begin{theorem}\label{theorem1}
Let $q$ be a prime power, and let $m$ and $\ell$ be integers satisfying
$3\leq\ell<m-1$ and
$\lfloor m/(\ell+1)\rfloor<\lfloor(m-1)/\ell\rfloor$.
Let $i$ and $s$ be integers satisfying
$\lfloor m/(\ell+1)\rfloor\leq i\leq s<\lfloor(m-1)/\ell\rfloor$,
and set $\delta=q^m-q^{m-1}-q^{m-1-i}-q^s-1$.
Then
\begin{equation}\label{equation2}
d_B\bigl(\C_{(q,m,\delta)}\bigr)
=\delta
\quad\hbox{and}\quad
d\bigl(\C_{(q,m,\delta)}\bigr)
=\delta+q^s,
\end{equation}
and
\begin{equation}\label{equation3}
\dim\bigl(\C_{(q,m,\delta)}\bigr)
=1+m(s-i+2)+\frac{m}{2}(m-2i+1)
+\sum_{k=3}^{\ell}\frac{m}{k}\binom{m-ki-1}{k-1}.
\end{equation}
\end{theorem}

The gap between the minimum distance and the Bose distance for the
family in Theorem~\ref{theorem1} is exactly $q^s$. For every fixed
 integer $\ell\geq3$, this gap grows without bound
as $m$ increases. Moreover, taking $\ell=3$ and $i=\lfloor m/4\rfloor$ in
Theorem~\ref{theorem1} recovers the family of \cite{ZhengEtAl2026} for
$m\geq13$ and shows that the minimum-distance lower bound established
there is attained for every prime power $q$.

To prove Theorem~\ref{theorem1}, we first establish the lower bound
using an argument similar to that in \cite{ZhengEtAl2026}.
Specifically, we show that
$\delta=q^m-q^{m-1}-q^{m-1-i}-q^s-1$ is a coset leader, and hence
$d_B(\C_{(q,m,\delta)})=\delta$. We then prove that $\C_{(q,m,\delta)}$ is contained in the
punctured generalized Reed--Muller code $\PGRM_q(\ell,m)$
\cite{DelsarteGoethalsMacWilliams1970}. 
The restrictions on its codeword weights derived from Ax's theorem
\cite{Ax1964}, together with the BCH bound, exclude all nonzero
codeword weights  from  $\delta$ to $\delta+q^s-1$, yielding
$d(\C_{(q,m,\delta)})\geq\delta+q^s$. It therefore remains to show
that $\C_{(q,m,\delta)}$ contains a codeword of weight
$\delta+q^s$. By the containment
$\C_{(q,m,\delta+q^s)}\subseteq\C_{(q,m,\delta)}$, this follows once
we prove that   $\C_{(q,m,\delta+q^s)}$ attains its
designed distance. Motivated by this observation, we show in the following theorem
that $\C_{(q,m,\delta+q^s)}$ attains its designed distance for
$1\leq i\leq\lfloor m/2\rfloor$.

\begin{theorem}\label{theorem2}
Let $q$ be a prime power, and let $m$ and $i$ be integers satisfying
$m\geq4$ and $1\leq i\leq\lfloor m/2\rfloor$. Set
$\delta=q^m-q^{m-1}-q^{m-1-i}-1$. Then
\begin{equation}\notag
d\bigl(\C_{(q,m,\delta)}\bigr)=\delta
\end{equation}
and
\begin{equation}\label{equation4}
\dim\bigl(\C_{(q,m,\delta)}\bigr)
=1+m+\frac{m}{2}(m-2i+1)
+\sum_{k=3}^{\lfloor m/(i+1)\rfloor}
\frac{m}{k}\binom{m-ki-1}{k-1}.
\end{equation}
\end{theorem}

Theorem~\ref{theorem2} recovers the binary minimum-distance result of
Kasami and Lin \cite{KasamiLin1972} and extends Li's nonbinary result
\cite[Theorem~1.2]{Li2017} to $1\leq i<\lfloor m/3\rfloor$.
For the proof, we use quadratic
forms to construct polynomials satisfying suitable conditions,
from which we obtain an $\F_q$-good zero-set of cardinality
$q^m-q^{m-1}-q^{m-1-i}$. By the criterion in
\cite[Theorem~1]{zhenggood}, this set yields a codeword of the
required weight.

For both families, we derive the dimension formulas by
characterizing and counting the integers in the complements of
the defining sets of the corresponding BCH codes.
The counting argument is based on the $q$-adic expansions
of these integers.

The remainder of this paper is organized as follows.
Section~\ref{section2} recalls the necessary preliminaries on BCH
codes, cyclotomic cosets, $\F_q$-good zero-sets, and punctured
generalized Reed--Muller codes. Section~\ref{section3} collects the
results on bilinear forms, quadratic forms
used to construct $\F_q$-good zero-sets. 
Sections~\ref{section4} and~\ref{section5} study the
families with $d=d_B$ and $d>d_B$ and prove
Theorems~\ref{theorem2} and~\ref{theorem1}, respectively. Finally, we conclude the paper in Section~\ref{section6}.

\section{Preliminaries on\! BCH codes and \! punctured\! generalized Reed--Muller codes}\label{section2}

\subsection{Primitive narrow-sense BCH codes}
Let $\mathbb{F}_q^n$ denote the $n$-dimensional vector space over
$\mathbb{F}_q$. A $q$-ary linear code of length $n$ is a linear
subspace of $\mathbb{F}_q^n$. Such a code $\mathcal{C}$ is called
\emph{cyclic} if
$(c_{n-1},c_0,\ldots,c_{n-2})\in\mathcal{C}$ whenever
$(c_0,c_1,\ldots,c_{n-1})\in\mathcal{C}$.
We identify each vector $(c_0,c_1,\ldots,c_{n-1})\in\mathbb{F}_q^n$
with its polynomial representation
\begin{equation}\notag
    c_0+c_1x+\cdots+c_{n-1}x^{n-1}
    \in\mathbb{F}_q[x]/(x^n-1).
\end{equation}
Under this identification, cyclic codes correspond precisely to
ideals of the quotient ring $\mathbb{F}_q[x]/(x^n-1)$.
Since this quotient ring is a principal ideal ring, every cyclic
code $\mathcal{C}$ is generated by a unique monic divisor $g(x)$
of $x^n-1$. We write $\mathcal{C}=\langle g(x)\rangle$ and call
$g(x)$ the \emph{generator polynomial} of $\mathcal{C}$.

Recall that $n=q^m-1$, and let $\beta$ be a primitive element of
$\F_{q^m}$. For each integer $j$ with $0\leq j\leq n-1$, let
$m_j(x)$ denote the minimal polynomial of $\beta^j$ over $\F_q$.
Since $g(x)$ divides $x^n-1$, whose roots are the distinct elements
$1,\beta,\ldots,\beta^{n-1}$, it follows that $g(x)$ is a product
of distinct such minimal polynomials.
In particular, the $q$-ary primitive narrow-sense BCH code
of length $n$ with designed distance $\delta$ is the cyclic code
generated by
\begin{equation}\notag
    g_{(q,m,\delta)}(x)
    =\operatorname{lcm}\bigl(m_1(x),m_2(x),\ldots,m_{\delta-1}(x)\bigr),
\end{equation}
where $\operatorname{lcm}$ denotes the least common multiple of polynomials and $2\leq\delta\leq n$.

For a cyclic code $\mathcal{C}=\langle g(x)\rangle$ of length
$n$ over $\F_q$, the set
\begin{equation}\notag
    T(\mathcal{C})
    =\{i:0\leq i\leq n-1,\ g(\beta^i)=0\}
\end{equation}
is called the \emph{defining set} of $\mathcal{C}$ with respect
to $\beta$. The dimension
of $\mathcal{C}$ is given by
\begin{equation}\label{equation5}
    \dim(\mathcal{C})=n-\left|T(\mathcal{C})\right|,
\end{equation}
where $|\cdot|$ denotes the cardinality of a set.
Moreover, for two cyclic codes $\mathcal{C}$ and $\mathcal{C}'$
of length $n$ over $\F_q$, we have
\begin{equation}\label{equation6}
    \mathcal{C}\subseteq\mathcal{C}'
    \quad\text{if and only if}\quad
    T(\mathcal{C}')\subseteq T(\mathcal{C}).
\end{equation}

\subsection{Cyclotomic cosets}
For two   integers $a$ and $b$ with $b\geq 1$,  we denote by  $a\bmod b$   the least nonnegative integer congruent to $a$ modulo $b$. Equivalently,
\begin{equation*}
    a\bmod b = a  - b  \left\lfloor \frac{a}{b} \right\rfloor.
\end{equation*}
For each integer $a$ with $0\leq a\leq n-1$, the \emph{$q$-cyclotomic coset} of $a$ modulo $n$ is defined by
\begin{equation*}
    C_a=\{aq^k\bmod n:k= 0,1,2,\ldots,m-1\}.
\end{equation*}
The smallest integer in $C_a$   is called the  \emph{coset leader} of $C_a$, denoted by $\cl(a)$. 
It is well known that 
   the minimal polynomial $m_a(x)$ of $\beta^a$ over $\mathbb{F}_q$ can be expressed as 
\begin{equation}\notag
    m_a(x)=\prod\limits_{i\in C_a}(x-\beta^i).
\end{equation} 
As a result, the defining set of 
$\mathcal{C}_{(q,m,\delta)}$ with respect to $\beta$ is precisely 
\begin{equation}\label{equation7}
  T(\mathcal{C}_{(q,m,\delta)}) = \bigcup_{a=1}^{\delta-1} C_a.
\end{equation}
Notice that $$ \{a\in \mathbb{Z}: 1\leq a\leq n-1 \} \setminus\bigcup_{a=1}^{\delta-1} C_a= \{a\in \mathbb{Z}: 1\leq a\leq n-1, \cl(a)\geq \delta\}.$$   Thus, 
 \begin{equation*}
    | T(\mathcal{C}_{(q,m,\delta)})| = n-1-|\{a\in \mathbb{Z}: 1\leq a\leq n-1, \cl(a)\geq \delta\}|
 \end{equation*}
Combining this with  \eqref{equation5}, we obtain
\begin{equation}\label{equation8}
\mathrm{dim}(\mathcal{C}_{(q,m,\delta)}) = 1+|\{a\in \mathbb{Z}: 1\leq a\leq n-1, \cl(a)\geq \delta\}|.
\end{equation}
Observe  that  for any  integer  $\delta$  satisfying $2\leq \delta\leq  n-1$, 
$
\bigcup\limits_{a=1}^{\delta-1}C_a \neq  \bigcup\limits_{a=1}^{\delta}C_a$
 if and only if 
$\delta$ is a coset leader.
  Consequently,   we have   
\begin{equation}\label{equation9}
d_B\left(\mathcal{C}_{(q,m,\delta)}\right)=\delta \quad \hbox{ if and only if } \quad \delta=\cl(\delta). 
\end{equation}

To simplify the subsequent discussion of the cardinalities and coset leaders of $q$-cyclotomic cosets, we introduce further notation. 
Let $\mathbb{Z}$ denote the set of all integers. 
 Note that 
each  integer $a$ with $0\leq a\leq n-1$ can be uniquely represented by its \textit{$q$-adic expansion} as  $a=\sum_{\ell=0}^{m-1}a_\ell q^\ell$, where  $a_\ell\in \mathbb{Z}$ satisfies $0\leq a_\ell\leq q-1$
 for all  $0\leq \ell\leq m-1$. We define the 
 \emph{$q$-adic vector} of $a$ by 
\begin{equation*}
    [a]_q=(a_{m-1},a_{m-2},\ldots,a_1,a_0). 
\end{equation*}
The \emph{$q$-weight} of a nonnegative integer $a$, denoted by
$\wtq(a)$, is the sum of the digits in its $q$-adic expansion.
 For a positive integer $k$,   let $\mathbb{Z}_{\geq 0}^k$  denote the set of all length-$k$ vectors of nonnegative integers, i.e., 
\begin{equation}
    \mathbb{Z}_{\geq 0}^k = \{(a_{k-1}, a_{k-2},\ldots, a_0): a_j\in \mathbb{Z}\hbox{ and }
        a_j\geq 0\text{ for all }0\leq j\leq k-1\}.
\end{equation}

 We equip $\mathbb{Z}_{\geq 0}^k$ with the lexicographic order. More precisely, for two distinct vectors
$
\mathbf{u}=(u_{k-1},\ldots,u_0) $ and $ \mathbf{w}=(w_{k-1},\ldots,w_0)$
in $\mathbb{Z}_{\geq 0}^k$, let $j$ be the largest integer  such that $u_j \neq w_j$. We write $\mathbf{u}<\mathbf{w}$ if $u_j<w_j$, and $\mathbf{u}\leq\mathbf{w}$ if either $\mathbf{u}<\mathbf{w}$ or $\mathbf{u}=\mathbf{w}$.
 Define the operator $\sigma$ on $\mathbb{Z}_{\geq0}^k$  by  
\begin{equation*}
    \sigma(a_{k-1},a_{k-2},\ldots,a_0) = (a_{k-2},\ldots,a_{0},a_{k-1}).   
\end{equation*}
 Let $\sigma^0$ be the identity map. For each positive integer $i$, let
$\sigma^i$ denote the $i$-fold composition of $\sigma$. Since $\sigma$
is bijective, so is $\sigma^i$ for all $i\geq 0$.

 The following proposition follows directly from the definitions.
\begin{proposition}\label{proposition1}
Let $a$ and $b$ be two integers with $0\leq a,b\leq n-1$. Then the following
statements hold.
\begin{enumerate}
    \item  $a=b$ if and only if $[a]_q=[b]_q$, and $a<b$ if and
    only if $[a]_q<[b]_q$.
    \item 
    $
    [a q^i\bmod n]_q=\sigma^i\left([a]_q\right)
$  for all $0\leq i\leq m-1$.
    \item  $a=\cl(a)$  if and only if
    $
    \sigma^i\left([a]_q\right)\geq  [a]_q
   $ 
    for all $1\leq i\leq m-1$.
\end{enumerate}
\end{proposition}

\subsection{Good zero-sets}
Let $\F_{q^m}[x]$ denote the polynomial ring over $\F_{q^m}$. Let $\F_{q^m}^*$ and $\F_q^*$ denote the multiplicative groups of
$\F_{q^m}$ and $\F_q$, respectively. 
For a finite subset $\mathcal{A}$ of $\F_{q^m}$, let $P_{\mathcal{A}}(x)$ denote
the unique monic polynomial of least degree that vanishes at
every element of $\mathcal{A}$, namely,
\begin{equation}\notag
    P_{\mathcal{A}}(x)=\prod_{\alpha\in \mathcal{A}}(x-\alpha).
\end{equation}
The set $\mathcal{A}$ is called an \emph{$\F_q$-good zero-set} if
$0\in \mathcal{A}$ and there exists $\lambda\in\F_{q^m}^*$ such that
\begin{equation}\notag
    P_{\mathcal{A}}'(\alpha)\in\lambda\F_q^*
    \qquad\text{for all }\alpha\in \mathcal{A},
\end{equation}
where $P_{\mathcal{A}}'$ denotes the formal derivative of $P_{\mathcal{A}}$.

The following lemma gives a criterion, in terms of
$\F_q$-good zero-sets, for a primitive narrow-sense BCH code
to have minimum distance equal to its designed distance. 
\begin{lemma}[{\cite[Theorem~1]{zhenggood}}]
\label{lemma1}
Let $\delta$ be an integer satisfying $2\leq\delta\leq q^m-1$.
Then $d(\mathcal{C}_{(q,m,\delta)})=\delta$ if and only if
there exists an $\F_q$-good zero-set
$\mathcal{A}\subseteq\F_{q^m}$ with $|\mathcal{A}|=\delta+1$.
\end{lemma}

For a polynomial $f\in\F_{q^m}[x]$, let $N(f)$ denote
the number of its distinct zeros in $\F_{q^m}$, that is,
\begin{equation}\notag
    N(f)=\bigl|\{x\in\F_{q^m}:f(x)=0\}\bigr|.
\end{equation}
The following lemma gives a construction of $\F_q$-good zero-sets
from polynomials satisfying certain conditions.  It will be used to determine the minimum distance
of the BCH codes in Theorem~\ref{theorem2}.  
\begin{lemma}\label{lemma2}   
Let $f\in\F_{q^m}[x]$ be a polynomial with
$1\leq\deg(f)<q^m$. Suppose that
$f(\F_{q^m})\subseteq\F_q$ and $N(f)=\deg(f)$.
Then there exists an $\F_q$-good zero-set
$\mathcal{A}\subseteq\F_{q^m}$ with
$|\mathcal{A}|=q^m-\deg(f)$.
\end{lemma}
\begin{proof}
    Since $\mathrm{deg}(f)<q^m$, there exists  $u\in \mathbb{F}_{q^m}$ such that $f(u)\ne0$. By replacing 
     $f(x)$ with  $f(x+u)$, if necessary, we may assume that $f(0)\neq 0$.  Define two sets 
\begin{equation*}
\mathcal{A}=\{\alpha\in \mathbb{F}_{q^m}:f(\alpha)\neq 0\}\qquad\hbox{and}\qquad
\mathcal{B}=\{\alpha\in \mathbb{F}_{q^m}:f(\alpha)=0\}.
\end{equation*}
Let $c$ be the leading coefficient of $f$. Since $N(f)= \mathrm{deg}(f)$, we have 
 \begin{equation*}
    f(x)=c\prod_{\alpha\in \mathcal{B}}(x-\alpha)=cP_{\mathcal{B}}(x).
 \end{equation*} Notice that $\mathcal{A}\cap \mathcal{B}=\varnothing$ and 
$
    \mathcal{A}\cup \mathcal{B}= \mathbb{F}_{q^m}.
$ 
We obtain 
\begin{equation*}
|\mathcal{A}|=|\mathbb{F}_{q^m}|-|\mathcal{B}|=q^m-N(f)=q^m-\mathrm{deg}(f)
\end{equation*} 
and \begin{equation*} 
 P_{\mathcal{A}}(x)P_{\mathcal{B}}(x)=\prod_{\alpha\in \mathbb{F}_{q^m}}(x-\alpha)=x^{q^m}-x.
\end{equation*} 
Differentiating gives $P_{\mathcal{A}}'P_{\mathcal{B}}+P_{\mathcal{A}}P_{\mathcal{B}}'=-1$. 
For $\alpha\in\mathcal{A}$, we have
$P_{\mathcal{A}}(\alpha)=0$ and
$P_{\mathcal{B}}(\alpha)=f(\alpha)/c\neq0$, and hence
\begin{equation}\notag
P_{\mathcal{A}}'(\alpha)
=-\frac{1}{P_{\mathcal{B}}(\alpha)}
=-\frac{c}{f(\alpha)}.
\end{equation}
Together with the assumption $f(\mathbb{F}_{q^m})\subseteq\F_q$, this shows that \begin{equation*}
    P_{\mathcal{A}}'(\alpha)\in c \F_q^*\quad  \text{for all } \alpha\in \mathcal{A}. 
\end{equation*}
Recalling that $f(0)\neq 0$, we also have $0\in \mathcal{A}$. Therefore, $\mathcal{A}$ is an $\F_q$-good zero-set with  $|\mathcal{A}|=q^m-\mathrm{deg}(f)$.
This completes the proof.
\end{proof}

\subsection{Punctured generalized Reed--Muller codes} 
For a polynomial
$f\in\mathbb{F}_q[x_1,\ldots,x_m]$, let $\deg_{x_i}(f)$ denote the degree of $f$ with respect to $x_i$, and
let $\deg(f)$ denote its total degree.
The polynomial $f$ is called \emph{reduced} if
$
\deg_{x_i}(f)\leq q-1$
for all $1\leq i\leq  m$.
Let $\ell$ be an integer with $1\leq\ell<m(q-1)$. Denote by
$\mathcal{P}_q(\ell,m)$ the set of all reduced polynomials
$f\in\mathbb{F}_q[x_1,\ldots,x_m]$ satisfying $\deg(f)\leq\ell$.
The generalized Reed--Muller code of order $\ell$ is defined by
\begin{equation}\notag
\mathrm{GRM}_q(\ell,m)
=
\left\{
\bigl(f(\mathbf{x})\bigr)_{\mathbf{x}\in\mathbb{F}_q^m}:
f\in\mathcal{P}_q(\ell,m)
\right\}.
\end{equation}
The punctured generalized Reed--Muller code of order
$\ell$, denoted by $\mathrm{PGRM}_q(\ell,m)$, is obtained from
$\mathrm{GRM}_q(\ell,m)$ by deleting the coordinate indexed by
the zero vector $\mathbf{0}\in\mathbb{F}_q^m$. That is, 
\begin{equation}\notag
\mathrm{PGRM}_q(\ell,m)
=
\left\{
\bigl(f(\mathbf{x})\bigr)_{\mathbf{x}\in\mathbb{F}_q^m \setminus 
\{\mathbf{0}\}}:
f\in\mathcal{P}_q(\ell,m)
\right\}.
\end{equation}
By the standard characterization \cite{DelsarteGoethalsMacWilliams1970,DingLiXia2018},  the code  $\mathrm{PGRM}_q(\ell,m)$ can be identified with the cyclic code of length $n=q^m-1$ whose defining set with respect to $\beta$ is
\begin{equation}\label{equation10}
T\left(\mathrm{PGRM}_q(\ell,m)\right)
=
\left\{
a:
1\leq a\leq n-1,\ 
\wtq(a)<m(q-1)-\ell
\right\}.
\end{equation}
By applying Ax's theorem \cite{Ax1964}, the authors of \cite{ZhengEtAl2026} established the following result on the weights of codewords in punctured generalized Reed--Muller codes.
\begin{lemma}\label{lemma3}
Let $\ell$ be an integer satisfying $1\leq \ell<m(q-1)$. Then every codeword
$\mathbf{c}\in\PGRM_q(\ell,m)$ satisfies
\begin{equation}\label{equation11}
\wt(\mathbf{c})
\equiv 0
\pmod{q^{\left\lfloor (m-1)/\ell\right\rfloor}}
\quad\text{or}\quad
\wt(\mathbf{c})
\equiv -1
\pmod{q^{\left\lfloor (m-1)/\ell\right\rfloor}}.
\end{equation}
\end{lemma}

\section{Bilinear forms and quadratic forms  over finite fields}\label{section3}
In this section, we recall basic definitions and results concerning
symmetric bilinear forms and quadratic forms over finite fields,
including their ranks and the numbers of solutions to quadratic
equations. These results will be used in the next section to construct polynomials satisfying the hypotheses of
Lemma~\ref{lemma2}. The resulting $\F_q$-good zero-sets
will then be used to determine the minimum distance of the BCH
codes in Theorem~\ref{theorem2}.

Let $V$ be a finite-dimensional vector space over $\F_q$.
A symmetric bilinear form on $V$ is a map
$B:V\times V\to\F_q$ satisfying
\begin{enumerate}
    \item[(1)]
    $B(k_1x_1+k_2x_2,y)=k_1B(x_1,y)+k_2B(x_2,y)$, 
    \item[(2)]
    $B(x,y)=B(y,x)$
\end{enumerate}
for all $x_1,x_2,x,y\in V$ and $k_1,k_2\in\F_q$.
The \emph{radical} of $B$ is defined by
\begin{equation}\notag
    \operatorname{Rad}(B)
    =\{x\in V:B(x,y)=0\text{ for all }y\in V\}.
\end{equation}
It is a  subspace of $V$, and the rank of $B$ is defined by
\begin{equation}\notag
    \operatorname{rank}(B)
    =\dim V-\dim \operatorname{Rad}(B).
\end{equation}

A quadratic form on $V$ is a map $Q:V\to\F_q$ such that 
    $Q(tx)=t^2Q(x)$ for all $t\in\F_q$ and $x\in V$, and 
    the map $B_Q:V\times V\to\F_q$ defined by
    \begin{equation}\notag
        B_Q(x,y)=Q(x+y)-Q(x)-Q(y)
    \end{equation}
    is a symmetric bilinear form on $V$.
We call $B_Q$  the  symmetric bilinear form associated with $Q$.  The radical of $Q$ is defined by
\begin{equation}\notag
    \operatorname{Rad}(Q)
    =\{x\in\operatorname{Rad}(B_Q):Q(x)=0\}.
\end{equation}
It is also a  subspace of $V$, and the rank of $Q$
is defined by
\begin{equation}\notag
    \operatorname{rank}(Q)
    =\dim V-\dim \operatorname{Rad}(Q).
\end{equation}

It is clear from the above definitions that  
 $\operatorname{rank}(Q)\geq\operatorname{rank}(B_Q)$.
The following lemma gives a more precise relation between these ranks.

\begin{lemma}[{\cite[Lemma~3.6]{Li2017}}]
\label{lemma4}
Let $Q$ be a quadratic form on a finite-dimensional vector space
over $\F_q$, and let $B_Q$ be the bilinear form associated with $Q$. Then
\begin{equation}\notag
\begin{cases}
    \operatorname{rank}(Q)=\operatorname{rank}(B_Q)
    &\text{if }q\text{ is odd},\\
    \operatorname{rank}(B_Q)\leq\operatorname{rank}(Q)
    \leq\operatorname{rank}(B_Q)+1
    &\text{if }q\text{ is even}.
\end{cases}
\end{equation}
\end{lemma}

\noindent\textbf{Remark.}
For odd $q$, Li \cite{Li2017} defines the bilinear form
associated with $Q$ as
$\frac{1}{2}[Q(x+y)-Q(x)-Q(y)]$, which equals
$\frac{1}{2}B_Q(x,y)$ under our convention.
Since multiplication by a nonzero scalar preserves the
radical and rank of a bilinear form,
Lemma~\ref{lemma4} remains valid
with our definition of $B_Q$.

By choosing a basis of $V$, we can express a quadratic form $Q$
on $V$ as a homogeneous polynomial of degree two in the coordinates
of $x\!\in \! V$ with respect to this basis;  see \cite[p.~182]{Hou2018}. 
In particular, when $q$ is even and $Q$ has odd rank, a suitable
choice of basis yields the expression given in the following lemma;
see also \cite[Corollary~6.48]{Hou2018} and
\cite[Theorem~6.30]{LidlNiederreiter1997}.

\begin{lemma}[{\cite[Proposition~3.8(2)]{Li2017}}]
\label{lemma5}
Suppose that $q$ is even. Let $V$ be an $m$-dimensional vector
space over $\F_q$, and let $Q:V\to\F_q$ be a quadratic form
of rank $2s+1$, where $0\le s\le\lfloor(m-1)/2\rfloor$.
Then there exists a  basis $\{e_1,\ldots,e_m\}$ of $V$
such that, for every $x=\sum_{j=1}^{m}x_je_j$ with
$x_1,\ldots,x_m\in\F_q$, we have
\begin{equation}\notag
    Q(x)=\sum_{j=1}^{s}x_{2j-1}x_{2j}
         +x_{2s+1}^2.
\end{equation}
\end{lemma}

The following results on the numbers of solutions to quadratic
equations over finite fields are given in
\cite[Propositions~3.9 and~3.10]{Li2017}.
\begin{lemma}
\textnormal{\cite[Proposition 3.9]{Li2017}}\label{lemma6}
Suppose that $q$ is odd. Let $ V$  be a vector space of dimension $m$ over
$\F_q$, and let $Q:V\to\F_q$ be a quadratic form of rank $r$.
Then for all  $h\in\F_q$,
\begin{equation}\notag
 \bigl|\{x\in V:Q(x)=h\}\bigr|=
 \begin{cases}
 q^{m-1}+\varepsilon\eta(h)q^{m-(r+1)/2} &\hbox{if } r\text{ is odd},\\
 q^{m-1}+\varepsilon\upsilon(h)q^{m-(r+2)/2} &\hbox{if } r\text{ is even},
 \end{cases}
\end{equation}
where  $\varepsilon\in\{1,-1\}$  depends only on $Q$, 
$\upsilon(h)=q-1$ if $h=0$ and $\upsilon(h)=-1$ otherwise, and 
\begin{equation}\notag
    \eta(x)=
    \begin{cases}
    1 &\hbox{if } x\text{ is a nonzero square in }\F_q,\\
    -1 &\hbox{if } x\text{ is not a square in }\F_q,\\
    0 &\hbox{if } x=0.
    \end{cases}
\end{equation} 
\end{lemma}

\begin{lemma}
    \textnormal{\cite[Proposition 3.10]{Li2017}}
    \label{lemma7}
Suppose that $q$ is even. Let $V$ be an $m$-dimensional vector space over
$\F_q$, and let $Q:V\to\F_q$ be a quadratic form of rank $r$.
Then for all $h\in \F_q$, \begin{equation*}
  \bigl|\{x\in V:Q(x)=h\}\bigr|=   \begin{cases}
        q^{m-1}& \hbox{if } r \text{ is odd},\\
        q^{m-1}+\varepsilon\upsilon(h)q^{m-(r+2)/2}&  \hbox{if } r \text{ is even},
    \end{cases}
\end{equation*}
where $\varepsilon\in\{1,-1\}$ depends only on $Q$, and
$\upsilon(h)=q-1$ if $h=0$ and $\upsilon(h)=-1$ otherwise.
\end{lemma}

We now turn to the quadratic forms on $\F_{q^m}$ used in the
proof of Theorem~\ref{theorem2}. 
Let  $m\ge4$ be an integer.  For each integer $i$ with  $1\le i\le\lfloor m/2\rfloor$, we define 
 the set 
\[
 E_i=\bigl\{\{a,b\}:0\le a<b\le m-1, i\le b-a\le m-i\bigr\}
\]
and the following set of polynomials: 
\begin{equation}\label{equation12}
 \mathcal{Q}_i=\left\{Q(x)=\sum_{\{a,b\}\in E_i}c_{\{a,b\}}x^{q^a+q^b}:
 c_{\{a,b\}}\in \F_{q^m}\text{ and } Q(\F_{q^m})\subseteq\F_q\right\}.
\end{equation}

We record two properties of $\mathcal{Q}_i$ that will be used to
construct $\F_q$-good zero-sets.
\begin{lemma}\label{lemma8}
Let $m$ and $i$ be integers satisfying $m\ge4$ and $1\le i\le\lfloor m/2\rfloor$.  Then each polynomial   
 $Q\in\mathcal{Q}_i$ satisfies  
\begin{equation}\notag
\deg Q\leq q^{m-1}+q^{m-1-i}.
\end{equation}
\end{lemma}

\begin{proof}
For every $\{a,b\}\in E_i$ with $a<b$, we have $b\leq m-1$
and $a\leq b-i\leq m-1-i$. Thus every monomial in
\eqref{equation12} has degree at most $q^{m-1}+q^{m-1-i}$,
which proves the assertion.
\end{proof}

The following lemma is a consequence of the rank results of Li
\cite[Propositions~4.6 and~5.2]{Li2017}.
For completeness, we give a proof in Appendix~\ref{appendix1}.

\begin{lemma}\label{lemma9}
There exists a quadratic form $Q\in\mathcal{Q}_i$ such that
\begin{enumerate}
 \item[(i)] if $q$ and $m$ are odd, then $\operatorname{rank}(Q)=2i+1$;
 \item[(ii)] if $q$ is odd and $m$ is even, then $\operatorname{rank}(Q)=2i$;
 \item[(iii)] if $q$ is even, then $\operatorname{rank}(B_Q)=2i$.
\end{enumerate}
\end{lemma}

Finally, we recall an auxiliary result on $\F_q$-linear maps.   A map $L:\F_{q^m}\to\F_{q^m}$ is called
$\F_q$-linear if
\begin{equation}\notag
L(ax+by)=aL(x)+bL(y)
\end{equation}
for all $a,b\in\F_q$ and $x,y\in\F_{q^m}$.
The following lemma gives a representation of such maps by
linearized polynomials.
\begin{lemma}\label{lemma10}
\textnormal{\cite[Proposition~2.28]{Hou2018}}
Every $\F_q$-linear map from $\F_{q^m}$ to $\F_{q^m}$ has a
polynomial representative of the form $\sum_{r=0}^{m-1}b_rx^{q^r}$,
where $b_r\in\F_{q^m}$.
\end{lemma}

\section{The family with $d=d_B$}\label{section4}
In this section, we prove Theorem~\ref{theorem2}, which determines the exact minimum distance and dimension of $\C_{(q,m,\delta)}$ for $\delta=q^m-q^{m-1}-q^{m-1-i}-1$ and provides the upper bound needed in the proof of Theorem~\ref{theorem1}.

\begin{proof}[Proof of Theorem~\ref{theorem2}]
We first prove that $d\bigl(\C_{(q,m,\delta)}\bigr)=\delta$.
It suffices to construct a polynomial $f\in\F_{q^m}[x]$ satisfying
\begin{equation}\label{equation13}
    f(\F_{q^m})\subseteq\F_q
    \quad\text{and}\quad
    \deg f=N(f)=q^{m-1}+q^{m-i-1}.
\end{equation}
Indeed, by Lemma~\ref{lemma2}, such a polynomial yields an
$\F_q$-good zero-set $\mathcal{A}\subseteq\F_{q^m}$ with
\begin{equation}\notag
    |\mathcal{A}|=q^m-q^{m-1}-q^{m-i-1}=\delta+1.
\end{equation}
By Lemma~\ref{lemma1},  this gives
$d\bigl(\C_{(q,m,\delta)}\bigr)=\delta$.
We now construct such  a polynomial $f$ by considering the following cases.

\textit{Case 1.} Suppose that $q$ and $m$ are odd.
By Lemma~\ref{lemma9}, we may choose
$Q\in\mathcal{Q}_i$ with $\operatorname{rank}(Q)=2i+1$.
By Lemma~\ref{lemma6}, we have
\begin{equation}\notag
    \bigl|\{x\in\F_{q^m}:Q(x)=h\}\bigr|
    =q^{m-1}+\varepsilon\eta(h)q^{m-i-1}
    \quad\text{for all }h\in\F_q,
\end{equation}
where $\varepsilon\in\{1,-1\}$ depends only on $Q$.
Since $q$ is odd, $\eta$ takes both values $1$ and $-1$ on $\F_q^*$.
Therefore, we may choose $h\in\F_q^*$ such that $\eta(h)=\varepsilon$.
Set $f=Q-h$. Then
\begin{equation}\notag
    N(f)=\bigl|\{x\in\F_{q^m}:Q(x)=h\}\bigr|
        =q^{m-1}+q^{m-i-1}.
\end{equation}
By the definition of $\mathcal{Q}_i$, we have
$f(\F_{q^m})\subseteq\F_q$.
Moreover, $f(0)=-h\ne0$, so $f$ is nonzero and
$N(f)\leq\deg f$.
By Lemma~\ref{lemma8}, we therefore have
\begin{equation}\notag
    N(f)\leq\deg f=\deg Q
    \leq q^{m-1}+q^{m-i-1}=N(f).
\end{equation}
Thus, $f$ satisfies \eqref{equation13}.

\medskip
\textit{Case 2.} Suppose that $q$ is odd and $m$ is even.
By Lemma~\ref{lemma9}, we may choose
$Q\in\mathcal{Q}_i$ with $\operatorname{rank}(Q)=2i$.
By Lemma~\ref{lemma6}, we have
\begin{equation}\label{equation14}
    \bigl|\{x\in\F_{q^m}:Q(x)=h\}\bigr|
    =q^{m-1}+\varepsilon\upsilon(h)q^{m-i-1}
    \qquad\text{for all }h\in\F_q,
\end{equation}
where $\varepsilon\in\{1,-1\}$ depends only on $Q$.
We claim that $\varepsilon=-1$.
Suppose, to the contrary, that $\varepsilon=1$.
Taking $h=0$ in \eqref{equation14} and using $q\geq3$, we obtain
\begin{equation}\notag
    N(Q)=q^{m-1}+(q-1)q^{m-i-1}
        >q^{m-1}+q^{m-i-1}.
\end{equation}
Since $\operatorname{rank}(Q)=2i>0$, the polynomial $Q$ is nonzero.
Hence $N(Q)\leq\deg Q$, contradicting the bound
$\deg Q\leq q^{m-1}+q^{m-i-1}$ in Lemma~\ref{lemma8}.
Therefore, $\varepsilon=-1$.

Now choose any $h\in\F_q^*$ and set $f=Q-h$.
Noting that $\upsilon(h)=-1$ for $h\in \F_q^*$, it follows from \eqref{equation14} that
\begin{equation}\notag
    N(f)=\bigl|\{x\in\F_{q^m}:Q(x)=h\}\bigr|
        =q^{m-1}+q^{m-i-1}.
\end{equation}
By the definition of $\mathcal{Q}_i$, we have
$f(\F_{q^m})\subseteq\F_q$.
Moreover, $f(0)=-h\ne0$, so $f$ is nonzero.
As in Case 1, we obtain  $\deg f=N(f)$, and hence $f$ satisfies \eqref{equation13}.

\medskip
\textit{Case 3.} Suppose that $q$ is even.
By Lemma~\ref{lemma9}, we may choose
$Q\in\mathcal{Q}_i$ with $\operatorname{rank}(B_Q)=2i$.
It follows from Lemma~\ref{lemma4} that
$\operatorname{rank}(Q)=2i$ or $2i+1$.
We consider these two subcases separately.

\smallskip
\textit{Subcase 3a.} Suppose that $\operatorname{rank}(Q)=2i$.
By Lemma~\ref{lemma7}, we have
\begin{equation}\label{equation15}
    \bigl|\{x\in\F_{q^m}:Q(x)=h\}\bigr|
    =q^{m-1}+\varepsilon\upsilon(h)q^{m-i-1}
    \qquad\text{for all }h\in\F_q,
\end{equation}
where $\varepsilon\in\{1,-1\}$ depends only on $Q$.
If $\varepsilon=-1$, choose any $h\in\F_q^*$ and set $f=Q-h$.
Noticing that
 $\upsilon(h)=-1$, it follows from \eqref{equation15} that
\begin{equation}\notag
    N(f)=q^{m-1}+q^{m-i-1}.
\end{equation}
Moreover, $f(0)=-h\ne0$, so $f$ is nonzero.

If $\varepsilon=1$, taking $h=0$ in \eqref{equation15} gives
\begin{equation}\notag
    N(Q)=q^{m-1}+(q-1)q^{m-i-1}.
\end{equation}
If $q>2$, this contradicts the degree bound in
Lemma~\ref{lemma8}, as in Case 2.
Hence $q=2$.
Set $f=Q$. Then $f$ is nonzero since
$\operatorname{rank}(Q)=2i>0$, and
$N(f)=q^{m-1}+q^{m-i-1}$.
In both cases, $f(\F_{q^m})\subseteq\F_q$.
As in Case 1, we have $\deg f=N(f)$, and hence
$f$ satisfies \eqref{equation13}.

\smallskip
\textit{Subcase 3b.} Suppose that $\operatorname{rank}(Q)=2i+1$.
By Lemma~\ref{lemma5}, there exists
an $\F_q$-basis $\{e_1,\ldots,e_m\}$ of $\F_{q^m}$ such that,
for   $x=\sum_{j=1}^{m}x_je_j$ with
$x_1,\ldots,x_m\in\F_q$, we have
\begin{equation}\notag
    Q(x)=\sum_{j=1}^{i}x_{2j-1}x_{2j}+x_{2i+1}^2.
\end{equation}
The coordinate map
$\sum_{j=1}^{m}x_je_j\mapsto x_{2i+1}$ is $\F_q$-linear.
By Lemma~\ref{lemma10}, this map can be represented by a polynomial
$L\in\F_{q^m}[x]$ with $\deg L\leq q^{m-1}$.
Set $f=Q+L$.
Then $f(\F_{q^m})\subseteq\F_q$, and by Lemma~\ref{lemma8},
\begin{equation}\notag
    \deg f\leq\max\{\deg Q,\deg L\}
    \leq q^{m-1}+q^{m-i-1}.
\end{equation}
In the above coordinates, we have
\begin{equation}\notag
    f(x)=\sum_{j=1}^{i}x_{2j-1}x_{2j}
         +x_{2i+1}^2+x_{2i+1}.
\end{equation}
Therefore,
\begin{equation}\notag
    N(f)=\left|\left\{
    (x_1,\ldots,x_m)\in\F_q^m:
    \sum_{j=1}^{i}x_{2j-1}x_{2j}
    +x_{2i+1}^2+x_{2i+1}=0
    \right\}\right|.
\end{equation}
We now count the vectors in the above set.
Fix a nonzero vector $(x_1,x_3,\ldots,x_{2i-1})\in\F_q^i$
and an element $x_{2i+1}\in\F_q$.
Choose $t$ with $1\leq t\leq i$ such that $x_{2t-1}\ne0$.
Then $(x_1,\ldots,x_m)$ belongs to the above set if and only if
\begin{equation}\notag
    x_{2t}
    =-
    \displaystyle\sum_{\substack{1\leq j\leq i\\j\ne t}} \frac{
    x_{2j-1}x_{2j}
    }{x_{2t-1}}+\frac{x_{2i+1}^2+x_{2i+1}}{x_{2t-1}}.
\end{equation}
Thus, the other $i-1$ coordinates among $x_2,x_4,\ldots,x_{2i}$
can be chosen arbitrarily, and each choice uniquely determines
$x_{2t}$.
This gives $q^{i-1}$ choices for $(x_2,x_4,\ldots,x_{2i})$,
and hence $(q^i-1)q\,q^{i-1}$ choices for the first
$2i+1$ coordinates.
If $(x_1,x_3,\ldots,x_{2i-1})=(0,\ldots,0)$,
then $(x_1,\ldots,x_m)$ belongs to the above set if and only if
$x_{2i+1}(x_{2i+1}+1)=0$, or equivalently,
$x_{2i+1}=0$ or $1$.
The coordinates $x_2,x_4,\ldots,x_{2i}$ can be chosen arbitrarily,
giving $2q^i$ choices for the first $2i+1$ coordinates.
In both cases, the remaining $m-2i-1$ coordinates are arbitrary.
Consequently,
\begin{equation}\notag
\begin{aligned}
    N(f)
    &=q^{m-2i-1}\bigl((q^i-1)q\,q^{i-1}+2q^i\bigr)\\
    &=q^{m-1}+q^{m-i-1}.
\end{aligned}
\end{equation}
Since $f(e_1+e_2)=1$, the polynomial $f$ is nonzero.
As in Case 1, we have $\deg f=N(f)$, and hence
$f$ satisfies \eqref{equation13}.

Thus, in all cases, there exists a polynomial satisfying \eqref{equation13}.
By Lemmas~\ref{lemma2} and~\ref{lemma1}, we obtain
$d\bigl(\C_{(q,m,\delta)}\bigr)=\delta$.

We now determine the dimension of $\C_{(q,m,\delta)}$. Let
\begin{equation}\notag
    \mathcal{S}
    =\{a\in\mathbb{Z}:1\leq a\leq n-1,\ \cl(a)\geq\delta\}.
\end{equation}
By \eqref{equation5}, we have
\begin{equation}\label{equation16}
    \dim(\C_{(q,m,\delta)})=1+|\mathcal{S}|.
\end{equation}
 Note that
\begin{equation}\label{equation17}
    [\delta]_q
    =\bigl(q-2,(q-1)^{(i-1)},q-2,(q-1)^{(m-i-1)}\bigr),
\end{equation}
where $x^{(r)}$ denotes a block consisting of $r $ consecutive copies of $x.$
For every $a\in\mathcal{S}$, all digits of $[a]_q$ belong to
$\{q-2,q-1\}$. Indeed, if a digit were at most $q-3$, then
a cyclic shift beginning with that digit would be smaller
than $[\delta]_q$, contradicting $\cl(a)\geq\delta$.
Moreover, since $a\leq n-1$, at least one digit of $[a]_q$
is equal to $q-2$.
For each integer $k$ with $1\leq k\leq m$, let $N_k$ denote
the number of integers $a\in\mathcal{S}$ for which $[a]_q$
contains exactly $k$ digits equal to $q-2$. 

Suppose that $[a]_q$ contains exactly  $k=1$ digit equal to $q-2$. Then there are $m$ possible positions for the unique  digit. Each resulting vector satisfies
$[\cl(a)]_q\geq[\delta]_q$. Hence
\begin{equation}\label{equation18}
    N_1=m.
\end{equation}

Suppose that $[a]_q$ contains exactly $k=2$ digits equal to $q-2$.
By \eqref{equation17}, we have $a\in\mathcal S$ if and only if
the two occurrences of $q-2$ in $[a]_q$ are separated by at
least $i-1$ digits equal to $q-1$ in each cyclic direction.
To count such integers $a$, we first choose a position in
$[a]_q$ at which to place a digit equal to $q-2$.
There are $m$ choices for this position. Once it is fixed,
the other digit equal to $q-2$ can be placed in
$m-1-2(i-1)=m-2i+1$ possible positions.
Each integer $a$ is counted twice, since either of the two
positions occupied by $q-2$ in $[a]_q$ can be chosen first.
Therefore,
\begin{equation}\label{equation19}
    N_2=\frac{m}{2}(m-2i+1).
\end{equation} 

Suppose that $[a]_q$ contains exactly $k\geq 3$ digits equal to $q-2$.
Choose a position occupied by $q-2$  in $[a]_q$ as the starting position,
and read  $[a]_q$ cyclically from that position.
Let $g_1,\ldots,g_k$ denote the numbers of digits equal to
$q-1$ between successive occurrences of $q-2$. Then
\begin{equation}\label{equation20}
    \sum\limits_{j=1}^kg_j=m-k.
\end{equation}

We claim that $a\in\mathcal{S}$ if and only if $g_j\geq i$
for all $1\leq j\leq k$. 
If $g_j\geq i$ for all such  $j$, then every cyclic shift of
$[a]_q$ beginning with $q-2$ is larger than $[\delta]_q$. Moreover, it is clear that
each cyclic shift  of $[a]_q$ beginning with $q-1$ is also larger
than $[\delta]_q$. Thus, $\cl(a)\geq\delta$ and 
$a\in\mathcal{S}$. 
Conversely, suppose that $a\in\mathcal{S}$. Then each cyclic shift of $[a]_q$ is greater than $[\delta]_q$. 
Comparing each cyclic shift of $[a]_q$ beginning with $q-2$ with $[\delta]_q$ in 
\eqref{equation17}, we first obtain $g_j\geq i-1$ for all $j$. 
Suppose that $g_{j_0}=i-1$ for some $j_0$.
Then by \eqref{equation20},   
\begin{equation}\notag
    g_{j}
    \leq m-k-g_{j_0}
    \leq m-k-(i-1)
    \leq m-i-2
\end{equation}
for all $1\leq j\leq k$ with $j\neq j_0$.
Consider the cyclic shift of $[a]_q$ beginning with the block
$\bigl(q-2,(q-1)^{(g_{j_0})}\bigr)$. 
This shift agrees with $[\delta]_q$ from the first digit through the second
occurrence of $q-2$. Starting from this second occurrence, the corresponding
blocks in this shift and in $[\delta]_q$ are
$\bigl(q-2,(q-1)^{(g_j)}\bigr)$ and
$\bigl(q-2,(q-1)^{(m-i-1)}\bigr)$, respectively,
where $j\neq j_0$ and $g_j\leq m-i-2$. Therefore, this shift is smaller than $[\delta]_q$,
leading to a contradiction.  
Thus, $g_j\geq i$ for all $j$, which proves the claim.

For a fixed starting position occupied by $q-2$, setting
$x_j=g_j-i$ gives a bijection between the admissible choices
of the remaining positions occupied by $q-2$ and the nonnegative integer
solutions to 
\begin{equation}\notag
    x_1+\cdots+x_k=m-k(i+1).
\end{equation}
There are $\binom{m-ki-1}{k-1}$ such solutions when
$k\leq\lfloor m/(i+1)\rfloor$, and none otherwise.
There are $m$ choices for the starting position. Each
vector is counted exactly $k$ times, since any of its $k$
positions occupied by $q-2$ can be chosen as the starting
position. 
Hence,
\begin{equation}\label{equation21}
    N_k=\frac{m}{k}\binom{m-ki-1}{k-1}
    \quad\text{for }
    3\leq k\leq\left\lfloor\frac{m}{i+1}\right\rfloor,
\end{equation}
and $N_k=0$ for all other $k\geq 3$.

Combining \eqref{equation18}, \eqref{equation19}, and
\eqref{equation21}, we obtain
\begin{equation}\notag
   |\mathcal{S}| = N_1+N_2+\sum\limits_{k=3}^{\lfloor m/(i+1)\rfloor}N_k
    =m+\frac{m}{2}(m-2i+1)
    +\sum_{k=3}^{\lfloor m/(i+1)\rfloor}
    \frac{m}{k}\binom{m-ki-1}{k-1},  
\end{equation}
where an empty sum is understood to be zero.
Together with \eqref{equation16}, this gives 
\eqref{equation4},  
 completing  the proof. 
\end{proof}

A $q$-ary linear code is called optimal if no $q$-ary linear code with the same length and dimension has a larger minimum distance. Table~\ref{table1} lists selected examples from Theorem~\ref{theorem2}, whose dimensions, Bose distances, and minimum distances have all been verified via Magma. By comparison with the best known lower bounds in Grassl's tables (\url{https://www.codetables.de}), every code in the table is either optimal or attains the best-known minimum distance.

\begin{table}[!ht]
\centering
\caption{Examples from Theorem~\ref{theorem2} with $d_B=d$.}
\label{table1}
\small
\setlength{\tabcolsep}{8pt}
\begin{tabular}{@{}ccccclc@{}}
\toprule
$q$ & $m$ & $n$ & $\mathrm{dim}$ & $d$ & Optimality & $i$\\
\midrule
$2$ & $4$ & $15$ & $7$ & $5$ & Optimal & $2$\\
$2$ & $5$ & $31$ & $11$ & $11$ & Optimal & $2$\\
$2$ & $6$ & $63$ & $16$ & $23$ & Best known & $2$\\
$2$ & $6$ & $63$ & $10$ & $27$ & Best known & $3$\\
\midrule
$3$ & $4$ & $80$ & $7$ & $50$ & Optimal & $2$\\
$4$ & $4$ & $255$ & $7$ & $187$ & Best known & $2$\\
\bottomrule
\end{tabular}
\end{table}

\section{The family with $d>d_B$}\label{section5}
Building on Theorem 
\ref{theorem2} and the weight divisibility of punctured generalized Reed--Muller codes in Lemma~\ref{lemma3}, we now prove Theorem \ref{theorem1}. We first record an elementary inequality whose proof is given in Appendix~\ref{appendix2}. 

\begin{lemma}\label{lemma11}
Suppose that $m$ and $\ell$ are integers satisfying
$
3\leq \ell<m-1
$
and $(m,\ell)\neq(7,3)$. Then
\begin{equation}\label{equation22} 
(\ell-1)\left\lfloor \frac{m}{\ell+1}\right\rfloor- \left\lfloor \frac{m-1}{\ell}\right\rfloor>0
\end{equation}
\end{lemma}

\begin{proof}[Proof of Theorem~\ref{theorem1}]
We divide the proof into four steps. We first show that
$d_B\bigl(\C_{(q,m,\delta)}\bigr)=\delta$ and then prove that
$\C_{(q,m,\delta)}\subseteq\PGRM_q(\ell,m)$.
In Step~3, we use this containment, together with
Lemma~\ref{lemma3} and the BCH bound, to obtain
$d\bigl(\C_{(q,m,\delta)}\bigr)\geq\delta+q^s$.
Combining this lower bound with the matching upper bound
obtained from Theorem~\ref{theorem2}, we conclude that
$d\bigl(\C_{(q,m,\delta)}\bigr)=\delta+q^s$.
Finally, we determine the dimension of $\C_{(q,m,\delta)}$.

\textbf{Step 1.} We first show that
$d_B\bigl(\C_{(q,m,\delta)}\bigr)=\delta$.
By \eqref{equation9}, it suffices to show that $\cl(\delta)=\delta$.
Set $b=m-i-s-2$. Then the $q$-adic vector of $\delta$ is
\begin{equation}\label{equation23}
    [\delta]_q
    =\bigl(q-2,(q-1)^{(i-1)},q-2,(q-1)^{(b)},q-2,
    (q-1)^{(s)}\bigr).
\end{equation}
Since $\ell\geq3$, we have
$m\geq3\left\lfloor(m-1)/\ell\right\rfloor$.
Together with $i\leq s<\left\lfloor(m-1)/\ell\right\rfloor$,
this gives
\begin{equation}\notag
\begin{aligned}
    b&=m-i-s-2\\
    &\geq m-2\left\lfloor\frac{m-1}{\ell}\right\rfloor\\
    &\geq\left\lfloor\frac{m-1}{\ell}\right\rfloor>s,
\end{aligned}
\end{equation}
and hence \begin{equation}\label{equation24}
i-1<s<b.\end{equation}

Every cyclic shift of $[\delta]_q$ beginning with $q-1$ is
larger than $[\delta]_q$. The shifts beginning with $q-2$
are $[\delta]_q$ itself and the following two vectors:
\begin{equation}\notag
    \sigma^i([\delta]_q)
    =\bigl(q-2,(q-1)^{(b)},q-2,(q-1)^{(s)},q-2,
    (q-1)^{(i-1)}\bigr)
\end{equation}
and
\begin{equation}\notag
    \sigma^{m-s-1}([\delta]_q)
    =\bigl(q-2,(q-1)^{(s)},q-2,(q-1)^{(i-1)},q-2,
    (q-1)^{(b)}\bigr).
\end{equation}
The inequality in (\ref{equation24}) implies  that both vectors are larger than $[\delta]_q$. 
Thus
\begin{equation}\notag
    [\delta]_q\leq\sigma^j([\delta]_q)
    \quad\text{for all }0\leq j\leq m-1.
\end{equation}
It follows that $\cl(\delta)=\delta$, and hence
$d_B\bigl(\C_{(q,m,\delta)}\bigr)=\delta$.

\textbf{Step 2.} We now prove that 
\begin{equation}\label{equation25}
 \C_{(q,m,\delta)}
 \subseteq\PGRM_q(\ell,m).
\end{equation}
By \eqref{equation6}, this is equivalent to 
\begin{equation}\notag
    T(\PGRM_q(\ell,m))\subseteq 
T(\C_{(q,m,\delta)}). 
\end{equation}

By \eqref{equation7}, 
  it is therefore  sufficient to show that 
  \begin{equation}\label{equation26}
      1\leq \cl(a)< \delta \quad \hbox{for all }a\in   T(\PGRM_q(\ell,m)) .
  \end{equation}

Let $a\in T(\PGRM_q(\ell,m))$. Then by \eqref{equation10}, we have \begin{equation}\notag
    1\leq a\leq n-1\quad \hbox{and} \quad \wt_q(a)<m(q-1)-\ell.
\end{equation}  Therefore, $\cl(a)\geq 1$, and we need only show that $\cl(a)<\delta$. 
If one digit of $[a]_q$ is at most $q-3$, then the first digit of
$[\cl(a)]_q$ is at most $q-3$, and hence
$\cl(a)<\delta$. 

We now assume that each digit of $[a]_q$ belongs to $\{q-2,q-1\}$.  Let 
$
    k$ denote the number of digits of $[a]_q$ equal to $q-2$. 
Then $[\cl(a)]_q$ has the form  
\begin{equation*}
    [\cl(a)]_q = \left( q-2, (q-1)^{(g_{1})}, q-2, (q-1)^{(g_2)}, \dots, q-2, (q-1)^{(g_k)} \right),
\end{equation*}
where $(g_{1}, g_{2},\ldots, g_k)\in \mathbb{Z}_{\geq 0}^k$ is a vector satisfying $\sum_{j=1}^{k}g_j=m-k$. Moreover, we have 
 \begin{equation}\label{equation27}
 g_{1}=\min \{g_j:1\leq j\leq k \}.
 \end{equation}  

Indeed, if $g_{1}>g_j$ for some $j$, then the cyclic shift of $[a]_q$ beginning with the block $\bigl(q-2,(q-1)^{(g_j)}\bigr)$ is smaller than $[\cl(a)]_q$, contradicting Proposition~\ref{proposition1}(3). 

Notice that 
 \begin{equation*}
    \wt_q(a)=(m-k)(q-1)+k(q-2)=m(q-1)-k.
\end{equation*}
Since $\wt_q(a)<m(q-1)-\ell$, this yields
$k\geq \ell+1$.  Combining this with \eqref{equation27} and
$\sum_{j=1}^{k}g_j=m-k$ yields
\begin{equation}\label{equation28}
    g_{1}
    \leq\left\lfloor\frac{m-k}{k}\right\rfloor
    \leq\left\lfloor\frac{m-\ell-1}{\ell+1}\right\rfloor
    =\left\lfloor\frac{m}{\ell+1}\right\rfloor-1
    \leq i-1,
\end{equation}
where the last inequality follows from
$i\geq\lfloor m/(\ell+1)\rfloor$.
If $g_{1}<i-1$, comparing the first $i$ digits
of $[\cl(a)]_q$ and $[\delta]_q$ gives $\cl(a)<\delta$.

It remains to consider the case $g_{1}=i-1$. In this case, the inequalities  in
\eqref{equation28}  force $i=\left\lfloor \frac{m}{\ell+1}\right\rfloor$ and $g_{1}=\left\lfloor \frac{m}{\ell+1}\right\rfloor-1$. 
We next show that $g_{2}<b$. 
First, it follows from  $\sum_{j=1}^{k}g_j=m-k$ and \eqref{equation27} that 
\begin{equation}\label{equation29}
\begin{split}
 g_{2}  &\leq m-k-(k-1)g_{1}\\
 &=  m-k-(k-1)\left(\left\lfloor\frac{m}{\ell+1}\right\rfloor-1\right)\\
 &\leq m-1-\ell \left\lfloor\frac{m}{\ell+1}\right\rfloor.
 \end{split}
\end{equation}
    On the other hand, 
since  $s< \lfloor \frac{m-1}{\ell} \rfloor$, we obtain 
\begin{equation*}
\begin{split}
    b&=m-i-s-2 \\
    &\geq m-\left\lfloor \frac{m}{\ell+1}\right\rfloor-\lfloor \frac{m-1}{\ell} \rfloor-1.
    \end{split}
\end{equation*}
Combining these two estimates gives 
\begin{equation*}
    b-g_{2}\geq (\ell-1)\left\lfloor \frac{m}{\ell+1}\right\rfloor- \left\lfloor \frac{m-1}{\ell}\right\rfloor.
\end{equation*}
If $(m,\ell)\neq (7,3)$, then by Lemma \ref{lemma11}, we have \begin{equation}\notag
(\ell-1)\left\lfloor \frac{m}{\ell+1}\right\rfloor- \left\lfloor \frac{m-1}{\ell}\right\rfloor>0.
\end{equation} It follows that $b-g_{2}>0$, and hence  $g_{2}<b$.

Suppose that $(m,\ell)=(7,3)$. Then $i=s=1$, $b=3$,
and $k\geq4$. Moreover, \eqref{equation29} yields
$g_{2}\leq3$. We claim that $g_{2}\leq2$.
Suppose, to the contrary, that $g_{2}=3$.
Since $\sum_{j=1}^{k}g_j=7-k$ and $k\geq4$,
we must have $k=4$ and $g_1=g_3=g_4=0$. Therefore,
\begin{equation}\notag
    [\cl(a)]_q
    =\bigl(q-2,q-2,(q-1)^{(3)},q-2,q-2\bigr).
\end{equation}
However,
\begin{equation}\notag
    \sigma^5\bigl([\cl(a)]_q\bigr)
    =\bigl(q-2,q-2,q-2,q-2,(q-1)^{(3)}\bigr)
    <[\cl(a)]_q,
\end{equation}
contradicting the definition of a coset leader.
Thus $g_{2}\leq2<b$.
We have therefore established $g_{2}<b$ in either case.
Together with $g_{1}=i-1$, this implies that
$\cl(a)<\delta$.
Combining all the cases considered above, we obtain
\eqref{equation26}, and hence \eqref{equation25} holds.
This completes Step 2.

\textbf{Step 3.} We now determine the minimum distance of $\mathcal{C}_{(q,m,\delta)}$.
Suppose that $\mathbf{c}$ is a nonzero codeword of $\mathcal{C}_{(q,m,\delta)}.$ Then it follows from \eqref{equation25} that $\mathbf{c}\in \PGRM_q(\ell,m)$. 
By applying Lemma~\ref{lemma3}, we obtain
\begin{equation}\label{equation30}
    \wt(\mathbf{c}) \bmod q^{\lfloor\frac{m-1}{\ell}\rfloor} = 0 
    \quad \text{or} \quad 
    \wt(\mathbf{c}) \bmod q^{\lfloor\frac{m-1}{\ell}\rfloor} = q^{\lfloor\frac{m-1}{\ell}\rfloor}-1.
\end{equation}
By the BCH bound, we also have $\wt(\mathbf{c})\geq \delta$. 
We claim that $\wt(\mathbf{c})\geq \delta+q^s$. Suppose, to the contrary, that $\wt(\mathbf{c})\leq \delta+q^s-1$. 
The assumptions $\ell\geq 3$ and $i<\left\lfloor\frac{m-1}{\ell}\right\rfloor$ imply
\begin{equation*}
    m-1-i\geq m-\left\lfloor\frac{m-1}{\ell}\right\rfloor> \left\lfloor\frac{m-1}{\ell}\right\rfloor.
\end{equation*}
Since $s<\left\lfloor\frac{m-1}{\ell}\right\rfloor$, this further yields that
\begin{equation*}
    \delta \bmod q^{\lfloor\frac{m-1}{\ell}\rfloor} = q^{\lfloor\frac{m-1}{\ell}\rfloor}-q^s-1.
\end{equation*}
Together with the inequalities $\wt(\mathbf{c})\geq \delta$ and $\wt(\mathbf{c})\leq \delta+q^s-1$, this implies 
\begin{equation*}
   0 < q^{\lfloor\frac{m-1}{\ell}\rfloor}-q^s-1 \leq \wt(\mathbf{c}) \bmod q^{\lfloor\frac{m-1}{\ell}\rfloor} \leq q^{\lfloor\frac{m-1}{\ell}\rfloor}-2,
\end{equation*}
which contradicts \eqref{equation30}. Therefore, the claim holds, i.e., $\wt(\mathbf{c})\geq \delta+q^s.$  Consequently, we have 
\begin{equation*}   d(\mathcal{C}_{(q,m,\delta)})\geq \delta+q^s.
\end{equation*}

We now show that the lower bound on the minimum distance is attained.
It is straightforward to verify that  $$\delta+q^s=q^m-q^{m-1}-q^{m-1-i}-1.$$  Moreover, the assumptions on $i$,  $m$ and $ \ell$ imply that
$1\leq i\leq \lfloor m/2 \rfloor$ and $m\geq 4$. 
Therefore, we can apply  Theorem  \ref{theorem2} to obtain  
   $   d(\mathcal{C}_{(q,m,\delta+q^s)})=\delta+q^s.
$
Note that 
$\mathcal{C}_{(q,m,\delta+q^s)}\subseteq\mathcal{C}_{(q,m,\delta)}$.  It follows that  \begin{equation*} 
d(\mathcal{C}_{(q,m,\delta)})\leq  d\left(\mathcal{C}_{(q,m,\delta+q^s)}\right)= \delta+q^s.\end{equation*}
Combining the upper and lower bounds on the minimum distance established as above, we obtain 
\begin{equation}\notag   d(\mathcal{C}_{(q,m,\delta)})=\delta+q^s.
\end{equation} 

\textbf{Step 4.} Finally, we determine the dimension of
$\C_{(q,m,\delta)}$. Set
\begin{equation}\notag
\mathcal S=\{a\in\mathbb Z:1\leq a\leq n-1,\ \cl(a)\geq\delta\}.
\end{equation}
By \eqref{equation5}, we have
\begin{equation}\label{equation31}
\dim\bigl(\C_{(q,m,\delta)}\bigr)=1+|\mathcal S|.
\end{equation}

Let $a\in\mathcal S$. Recalling the $q$-adic expansion of $\delta$ in \eqref{equation23}, we can see that every digit of $[a]_q$ belongs to $\{q-2, q-1\}$. 
Since $a\leq n-1=q^m-2$,
the vector $[a]_q$ contains at least one digit equal to $q-2$.    Moreover, 
we claim that $[a]_q$ contains at most $\ell$ digits equal to
$q-2$. Suppose, to the contrary, that it contains exactly $k$
such digits for some  $k>\ell$. Then
\begin{equation}\notag
\wtq(a)=m(q-1)-k<m(q-1)-\ell.
\end{equation}
By \eqref{equation10}, this implies that $a\in T(\PGRM_q(\ell,m))$.
It follows from \eqref{equation26}, established in Step~2, that
$\cl(a)<\delta$, contradicting $a\in\mathcal S$.
Therefore, $[a]_q$ contains at most $\ell$ digits equal to $q-2$,
as claimed.
For each integer $k$ with  $1\leq k\leq\ell$, let $N_k$ denote  the number of integers
$a\in\mathcal S$ such that $[a]_q$ contains exactly $k$
digits equal to $q-2$.

Proceeding as in the proof of Theorem~\ref{theorem2}, we first obtain 
\begin{equation}\label{equation32}
N_1=m\quad\hbox{ and }\quad 
N_2=\frac{m}{2}(m-2i+1).
\end{equation}

Suppose that $[a]_q$ contains exactly $k=3$ digits equal to
$q-2$. 
Choose a position occupied by $q-2$ in $[a]_q$ as the
starting position, and let $g_1,g_2,g_3$ denote the numbers
of digits  between successive occurrences of
$q-2$, read cyclically from left to right starting at this
position. Then
\begin{equation}\label{equation33}
    g_1+g_2+g_3=m-3.
\end{equation}
We claim that $a\in\mathcal S$ if and only if either
$g_j\geq i$ for all $1\leq j\leq3$, or the starting
position in $[a]_q$ can be chosen so that
$
    g_1=i-1,  g_2\geq b,$ and $ g_3\geq i.
$

We first prove the sufficiency. If $g_j\geq i$ for all
$1\leq j\leq3$, then every cyclic shift of $[a]_q$ is
larger than $[\delta]_q$, and hence $a\in\mathcal S$.
Now suppose that the second condition holds.
Recall from \eqref{equation24} that $i\leq s<b$.
If $g_2>b$, then the cyclic shift beginning at the chosen
starting position is larger than $[\delta]_q$.
If $g_2=b$, then
\begin{equation}\notag
    g_3=m-3-(i-1)-b=s,
\end{equation}
so this shift is exactly $[\delta]_q$.
The shifts beginning at the other two positions occupied
by $q-2$ are larger than $[\delta]_q$, since
$g_2,g_3\geq i$. Every shift beginning with $q-1$ is also
larger than $[\delta]_q$. Thus,  we have $a\in\mathcal S$.

Conversely, suppose that $a\in\mathcal S$.
Comparing every cyclic shift of $[a]_q$ beginning with $q-2$ with $[\delta]_q$ in \eqref{equation23} gives $g_j\geq i-1$ for all $1\leq j\leq3$.

If $g_j\geq i$ for all $1\leq j\leq 3$ does not hold, then we can choose the starting position in $[a]_q$
so that $g_1=i-1$.
Comparing the shift beginning at this position with
$[\delta]_q$ gives $g_2\geq b$.
If $g_3=i-1$, then comparison of the shift beginning with
the block $\bigl(q-2,(q-1)^{(g_3)},q-2\bigr)$ with
$[\delta]_q$ would imply $g_1\geq b$, contradicting
$g_1=i-1<b$. Therefore, $g_3\geq i$, and the second
condition holds. This proves the claim.

We now count the integers satisfying these two conditions.
By the counting argument in the proof of
Theorem~\ref{theorem2}, the number of integers satisfying
$g_j\geq i$ for all $1\leq j\leq3$ is
\begin{equation}\notag
    \frac{m}{3}\binom{m-3i-1}{2}.
\end{equation}
For the second condition, once a starting position occupied
by $q-2$ in $[a]_q$ is fixed, the position of the second
occurrence of $q-2$ is determined, since $g_1=i-1$. Furthermore, by \eqref{equation33}, 
 the conditions $g_2\geq b$ and $g_3\geq i$ hold if and only if 
\begin{equation}\notag
    b\leq g_2\leq m-2i-2.
\end{equation}
Thus,  there are
\begin{equation}\notag
    m-2i-1-b=s-i+1
\end{equation}
choices for  the position of the third occurrence of $q-2$. 
Each choice of these three positions occupied by $q-2$
determines a unique integer $a$.
There are $m$ choices for the starting position in $[a]_q$. Since $g_1=i-1$ and $g_2,g_3\geq i$, each integer $a$ has
a unique starting position satisfying the second condition,  
and is therefore counted exactly once.
 Hence the second condition
contributes $m(s-i+1)$ integers.
Combining the two counts, we obtain
\begin{equation}\label{equation34}
    N_3=\frac{m}{3}\binom{m-3i-1}{2}+m(s-i+1).
\end{equation}

Finally, suppose that $[a]_q$ contains exactly $k$ digits
equal to $q-2$ for some  $4\leq k\leq \ell$. 
Choose an occurrence of $q-2$ in $[a]_q$ as the starting
point, and let $g_1,\ldots,g_k$ denote the numbers of digits
 between successive occurrences of $q-2$,
read cyclically from left to right starting at this
position.  Then
\begin{equation}\label{equation35}
    \sum_{j=1}^{k}g_j=m-k.
\end{equation}
We claim that $a\in\mathcal S$ if and only if $g_j\geq i$
for all $1\leq j\leq k$.
If every $g_j\geq i$, then every cyclic shift of $[a]_q$
is larger than $[\delta]_q$, and hence $a\in\mathcal S$.

Conversely, suppose that $a\in\mathcal S$. Then 
comparing  each cyclic shift of $[a]_q$ with $[\delta]_q$  
gives $g_j\geq i-1$ for all $1\leq j\leq k$.
Suppose  that $g_{j_0}=i-1$ for some $j_0$.
By choosing the starting occurrence of $q-2$ appropriately,
we may assume that $g_1=i-1$.
Then it  follows   that
\begin{equation}\notag
\begin{aligned}
    g_2
    \leq m-k-(k-1)(i-1)
    \leq m-3i-1
    \leq b,
\end{aligned}
\end{equation}
where the last inequality follows from $\lfloor m/(\ell+1)\rfloor \leq i\leq s<\lfloor(m-1)/\ell\rfloor$  and $b=m-i-s-2.$

On the other hand, comparing the cyclic shift of $[a]_q$ 
beginning with $\bigl(q-2,(q-1)^{(g_1)},q-2\bigr)$
with $[\delta]_q$ in 
\eqref{equation23} gives $g_2\geq b$. Thus, we have $g_2=b$. Moreover, when $g_2=b$, the same comparison gives  $g_3\geq s$.  It follows that  
\begin{equation}\notag
    g_1+g_2+g_3\geq(i-1)+b+s=m-3>m-k,
\end{equation}
contradicting \eqref{equation35}.  Therefore, $g_j\geq i$ for all $1\leq j\leq k$,
which proves the claim.

Applying an analogous  counting argument as in the proof of
Theorem~\ref{theorem2}, we obtain
\begin{equation}\label{equation36}
    N_k=\frac{m}{k}\binom{m-ki-1}{k-1}
    \qquad\text{for }4\leq k\leq\ell.
\end{equation}
Combining \eqref{equation32}, \eqref{equation34}, and
\eqref{equation36} with \eqref{equation31},
we obtain
\begin{equation}\notag
\dim\bigl(\C_{(q,m,\delta)}\bigr)
=1+m(s-i+2)+\frac{m}{2}(m-2i+1)
+\sum_{k=3}^{\ell}\frac{m}{k}\binom{m-ki-1}{k-1}.
\end{equation}
This completes the proof.
\end{proof}

Table~\ref{table2} lists several examples from
Theorem~\ref{theorem1}. The dimensions and Bose distances of all codes 
in the table have been verified using Magma. The minimum distance
of the code in the first row has also been verified using Magma.
For the remaining codes, we were unable to complete the
minimum-distance verifications in Magma within a reasonable
amount of time.

\begin{table}[ht]
\centering
\caption{Some binary and ternary primitive narrow-sense BCH codes
obtained from Theorem~\ref{theorem1}.}
\label{table2}
\small
\setlength{\tabcolsep}{4pt}
\begin{tabular}{@{}ccrrrrrc@{}}
\toprule
$q$ & $m$ & $n$ & $\mathrm{dim}$ & $d$ & $d_B$ & $d-d_B$
& $(\ell,i,s)$\\
\midrule
$2$ & $7$  & $127$  & $43$  & $31$   & $29$   & $2$
& $(3,1,1)$\\
$2$ & $9$  & $511$  & $94$  & $127$  & $125$  & $2$
& $(4,1,1)$\\
$2$ & $10$ & $1023$ & $66$  & $383$  & $379$  & $4$
& $(3,2,2)$\\
$2$ & $11$ & $2047$ & $89$  & $767$  & $763$  & $4$
& $(3,2,2)$\\
$2$ & $11$ & $2047$ & $221$ & $511$  & $509$  & $2$
& $(5,1,1)$\\
$2$ & $13$ & $8191$ & $92$  & $3583$ & $3575$ & $8$
& $(3,3,3)$\\
$2$ & $13$ & $8191$ & $170$ & $3071$ & $3067$ & $4$
& $(4,2,2)$\\
$2$ & $13$ & $8191$ & $547$ & $2047$ & $2045$ & $2$
& $(6,1,1)$\\
\midrule
$3$ & $7$  & $2186$    & $43$  & $1214$    & $1211$    & $3$
& $(3,1,1)$\\
$3$ & $9$  & $19682$   & $94$  & $10934$   & $10931$   & $3$
& $(4,1,1)$\\
$3$ & $10$ & $59048$   & $66$  & $37178$   & $37169$   & $9$
& $(3,2,2)$\\
$3$ & $11$ & $177146$  & $89$  & $111536$  & $111527$  & $9$
& $(3,2,2)$\\
$3$ & $11$ & $177146$  & $221$ & $98414$   & $98411$   & $3$
& $(5,1,1)$\\
$3$ & $13$ & $1594322$ & $92$  & $1043198$ & $1043171$ & $27$
& $(3,3,3)$\\
$3$ & $13$ & $1594322$ & $170$ & $1003832$ & $1003823$ & $9$
& $(4,2,2)$\\
$3$ & $13$ & $1594322$ & $547$ & $885734$  & $885731$  & $3$
& $(6,1,1)$\\
\bottomrule
\end{tabular}
\end{table}

\section{Conclusion}\label{section6}
In this paper, we have determined the exact minimum distances, Bose
distances, and dimensions of two infinite families of $q$-ary
primitive narrow-sense BCH codes for every prime power $q$.
Theorem~\ref{theorem1} establishes an infinite family satisfying
$d-d_B=q^s$, so that for every fixed $q$ and $\ell\geq3$, the gap
between the minimum distance and the Bose distance grows without bound
as $m$ increases. Theorem~\ref{theorem2} determines a closely related
family satisfying $d=d_B$ for $m\geq4$ and
$1\leq i\leq\lfloor m/2\rfloor$. Its proof uses quadratic forms and
$\F_q$-linear maps to construct $\F_q$-good zero-sets, which establish
the existence of minimum-weight codewords via the good zero-set
criterion and thereby yield the matching upper bound needed to prove
Theorem~\ref{theorem1}. It would be interesting to construct further
infinite families of primitive narrow-sense BCH codes with $d>d_B$ and
determine their exact minimum distances.

\section*{Acknowledgments}
The author gratefully acknowledges Professor Maosheng Xiong and Professor Cunsheng Ding for their financial support. The author acknowledge the
use of ChatGPT (sol 6.1 and Astra 6) for exploratory discussions during the early phase of the research. All mathematical results were developed,
proved, and independently verified by the author, who take full responsibility for the content and integrity of this work.
\appendix 

\section{Proof of Lemma \ref{lemma9}}\label{appendix1}
In this appendix, we recall the rank results of Li
\cite[Propositions~4.6 and~5.2]{Li2017} in their trace
formulation and explain how they yield Lemma~\ref{lemma9}
in the polynomial form used in this paper.

Let $m\geq4$ be an integer, and let $\Tr_{q}^{q^m}$ denote the
trace map from $\F_{q^m}$ to $\F_q$.
For each integer $j$ with $(m-2)/2\leq j\leq m-2$, let 
\begin{equation}\label{equation37}
\widetilde{\mathcal Q}_j
=\left\{
\Tr_{q}^{q^m}\left(\sum_{\ell=(m+1)/2}^{j+1}
\lambda_\ell x^{q^\ell+1}\right):
\lambda_\ell\in\F_{q^m}\right\}
\qquad\text{if }m\text{ is odd},
\end{equation}
and
\begin{equation}\label{equation38}
\begin{aligned}
\widetilde{\mathcal Q}_j
=\biggl\{&\Tr_{q}^{q^{m/2}}
\left(\lambda_{m/2}x^{q^{m/2}+1}\right)
+\Tr_{q}^{q^m}\left(\sum_{\ell=m/2+1}^{j+1}
\lambda_\ell x^{q^\ell+1}\right):\\
&\lambda_{m/2}\in\F_{q^{m/2}}, 
\lambda_\ell\in\F_{q^m}\text{ for } m/2+1\leq\ell\leq j+1
\biggr\}
\qquad\text{if }m\text{ is even}.
\end{aligned}
\end{equation}
In either case, $\widetilde{\mathcal Q}_j$ consists of quadratic
forms on $\F_{q^m}$. 

A set $\mathcal{S}$ of bilinear forms is called a $d$-code if
$\operatorname{rank}(B-C)\geq d$ for all distinct $B,C\in\mathcal{S}$.
It is called a \emph{proper} $d$-code if, in addition, equality
holds for some such pair.
The following lemma records the proper
code parameters established by Li 
\cite{Li2017}.

\begin{lemma}[{\cite[Propositions~4.6 and~5.2]{Li2017}}]
\label{lemma12}
Let $j$ be an integer with $(m-2)/2\leq j\leq m-2$, and let
$\widetilde{\mathcal Q}_j$ be defined by
\eqref{equation37} or
\eqref{equation38}, depending on whether $m$ is odd or even.
Then the set
\begin{equation}\notag
\mathcal{S}_j=\{B_Q:Q\in\widetilde{\mathcal Q}_j\}
\end{equation}
is a proper $d_j$-code, where
\begin{equation}\notag
d_j=
\begin{cases}
2m-2j-1 &\text{if }q\text{ and }m\text{ are odd},\\
2m-2j-2 &\text{otherwise}.
\end{cases}
\end{equation}
\end{lemma}

\begin{proof}[Proof of Lemma~\ref{lemma9}]
Let $i$ be an integer with  $1\leq i\leq\lfloor m/2\rfloor$, and set $j=m-i-1$.
Then $(m-2)/2\leq j\leq m-2$.
By Lemma~\ref{lemma12},
$\mathcal S_j$ is a proper $d_j$-code, where $d_j=2i+1$
if $q$ and $m$ are odd, and $d_j=2i$ otherwise.
Notice that $\mathcal S_j$ is closed under subtraction.
Thus, there exists
$Q\in\widetilde{\mathcal Q}_j$ such that
$\operatorname{rank}(B_Q)=d_j$.

We now show that $Q$ has a polynomial representative in $\mathcal Q_i$.
Suppose first that $m$ is odd.
Recalling \eqref{equation37}, we have
\begin{equation}\notag
Q(x)=
\sum_{\ell=(m+1)/2}^{j+1}
\sum_{r=0}^{m-1}
\lambda_\ell^{q^r}
x^{q^{(\ell+r)\bmod m}+q^r}.
\end{equation}
Since $j+1=m-i\leq m-1$, the summation indices satisfy
\begin{equation}\notag
\frac{m+1}{2}\leq\ell+r\leq2m-2.
\end{equation}
We then distinguish two cases.

\medskip
\textit{Case 1.} Suppose that $\ell+r\leq m-1$.
Then $(\ell+r)\bmod m=\ell+r$, and
$0\leq r<\ell+r\leq m-1$.
Moreover, since $i\leq \lfloor \frac{m}{2}\rfloor$ and $j=m-i-1$, we also have  
\begin{equation}\notag
i
\leq(\ell+r)-r
\leq m-i.
\end{equation}
It follows that $\{r,\ell+r\}\in E_i$.

\medskip
\textit{Case 2.} Suppose that $\ell+r\geq m$.
Then $(\ell+r)\bmod m=\ell+r-m$.
Noting that $\ell\leq j+1<m$, we have
\begin{equation}\notag
0\leq\ell+r-m<r\leq m-1.
\end{equation}
Furthermore,
\begin{equation}\notag
i\leq r-(\ell+r-m)=m-\ell
\leq\frac{m-1}{2}\leq m-i.
\end{equation}
Hence $\{\ell+r-m,r\}\in E_i$.

Now suppose that $m$ is even.
By \eqref{equation38},
we have
\begin{equation}\notag
\begin{aligned}
Q(x)
={}&\sum_{r=0}^{m/2-1}
\lambda_{m/2}^{q^r}x^{q^r+q^{r+m/2}}\\
&+\sum_{\ell=m/2+1}^{j+1}\sum_{r=0}^{m-1}
\lambda_\ell^{q^r}x^{q^{(\ell+r)\bmod m}+q^r}.
\end{aligned}
\end{equation}
For the first sum, $0\leq r<r+m/2\leq m-1$ and
$i\leq m/2\leq m-i$, so $\{r,r+m/2\}\in E_i$.
In the second sum, each index $\ell$ satisfies
$m/2+1\leq\ell\leq j+1=m-i$.
Thus, the same argument as in the two cases above shows that
every index pair in the second sum belongs to $E_i$.

In either case, collecting the terms gives a polynomial representative
of $Q$ whose index pairs all belong to $E_i$. Moreover, 
by the definition of the trace maps, we have 
 $Q(\F_{q^m})\subseteq \F_q$. Therefore, we conclude that $Q\in \mathcal Q_i$.

Finally, 
if  $q$ is odd, then Lemma~\ref{lemma4}
gives
\begin{equation}\notag
\operatorname{rank}(Q)=\operatorname{rank}(B_Q)=d_j,
\end{equation}
which proves parts~\textup{(i)} and~\textup{(ii)}.
If $q$ is even, we already have
$\operatorname{rank}(B_Q)=d_j=2i$, proving
part~\textup{(iii)}.
This completes the proof.
\end{proof}

\section{Proof of Lemma~\ref{lemma11}}\label{appendix2}

\begin{proof}
By the division algorithm, there exist integers $k$ and $r$ with $0\leq r\leq \ell$ such that 
\begin{equation}\notag
m=(\ell+1)k+r.
\end{equation}
This implies that $\lfloor\frac{m}{\ell+1}\rfloor=k$, and hence 
\begin{equation}\label{equation39}
\begin{split}
m-1&=(\ell+1)\left\lfloor\frac{m}{\ell+1}\right\rfloor+r-1\\
&\leq
(\ell+1)\left\lfloor\frac{m}{\ell+1}\right\rfloor+\ell-1.
\end{split}
\end{equation}

First, suppose that $\ell\geq4$. Since $\ell<m-1$, we have $
\left\lfloor\frac{m}{\ell+1}\right\rfloor\geq1.
$ Therefore, 
\begin{equation}\notag
\begin{split}
\ell(\ell-1)\left\lfloor\frac{m}{\ell+1}\right\rfloor
-\left(
(\ell+1)\left\lfloor\frac{m}{\ell+1}\right\rfloor+\ell-1
\right)&=
(\ell^2-2\ell-1)
\left\lfloor\frac{m}{\ell+1}\right\rfloor-(\ell-1)\\
&\geq
\ell^2-2\ell-1-(\ell-1)>0.
\end{split}
\end{equation}
Consequently, 
\begin{equation}\notag
\ell(\ell-1)\left\lfloor\frac{m}{\ell+1}\right\rfloor
>
(\ell+1)\left\lfloor\frac{m}{\ell+1}\right\rfloor+\ell-1.
\end{equation}
Combining this with \eqref{equation39} gives 
\begin{equation}\notag
  \ell(\ell-1)
\left\lfloor\frac{m}{\ell+1}\right\rfloor-(m-1)>0.  
\end{equation}
This implies that 
 \eqref{equation22} holds. 

It remains to consider the case $\ell=3$. 
Since $(m,\ell)\neq(7,3)$ and $\ell<m-1$, we have either $m=5,6$ or $m\geq8$. For $m=5,6$, the desired inequality in \eqref{equation22} follows by direct verification. If $m\geq8$, then
$
\left\lfloor\frac m4\right\rfloor >1,
$ 
and hence
\begin{equation}\notag
m-1
\leq
4\left\lfloor\frac m4\right\rfloor+2
<
6\left\lfloor\frac m4\right\rfloor.
\end{equation}
It follows that 
\begin{equation}\notag
2\left\lfloor\frac m4\right\rfloor-\left\lfloor\frac{m-1}{3}\right\rfloor>0.
\end{equation}
That is, \eqref{equation22} holds. This completes the proof. 
\end{proof}

\end{document}